\documentclass[review]{elsarticle}

\usepackage{adjustbox}
\usepackage{amsmath}
\usepackage{algorithm}
\usepackage{algorithmicx}
\usepackage{algpseudocode}
\usepackage[normalem]{ulem}

\journal{Journal of \LaTeX\ Templates}

\begin{document}

\begin{frontmatter}

\title{Exploiting Investors Social Network for Stock Prediction in China's Market}

\author[mymainaddress]{Xi Zhang\corref{mycorrespondingauthor}}
\cortext[mycorrespondingauthor]{Corresponding author}
\ead{zhangx@bupt.edu.cn}

\author[mymainaddress]{Jiawei Shi}
\ead{2011212788@bupt.edu.cn}

\author[mymainaddress]{Di Wang}
\ead{jxgzwd@bupt.edu.cn}

\author[mymainaddress,mysecondaddress]{Binxing Fang}
\ead{fangbx@bupt.edu.cn}

\address[mymainaddress]{Key Laboratory of Trustworthy Distributed Computing and Service, Ministry of Education, Beijing University of Posts and Telecommunications, Beijing 100876, China}

\address[mysecondaddress]{Institute of Electronic and Information Engineering of UESTC in Guangdong, Dongguan Guangdong 523808, China}

\begin{abstract}
Recent works have shown that social media platforms are able to influence the trends of stock price movements. However, existing works have majorly focused on the U.S. stock market and lacked attention to certain emerging countries such as China, where retail investors dominate the market. In this regard, as retail investors are prone to be influenced by news or other social media, psychological and behavioral features extracted from social media platforms are thought to well predict stock price movements in the China's market. Recent advances in the investor social network in China enables the extraction of such features from web-scale data. In this paper, on the basis of tweets from Xueqiu, a popular Chinese Twitter-like social platform specialized for investors, we analyze features with regard to collective sentiment and perception on stock relatedness and predict stock price movements by employing nonlinear models. The features of interest prove to be effective in our experiments.
\end{abstract}


\end{frontmatter}


\section{Introduction}

Social networks such as Twitter, Weibo, Facebook, and LinkedIn have attracted millions of users to post and acquire information, which have been well studied by various works~\cite{kwak2010twitter,su2016understanding,anderson2015global,viswanath2009evolution}. In addition to these general social networks, there is another breed of smaller, more focused sites that cater to niche audiences. Here we look at a social site designed for traders and investors, that is, Xueqiu. Xueqiu is a specialized social network for Chinese investors of the stock market, and due to the increasing number of retail investors, Xueqiu has attracted millions of users. Xueqiu enables investors to share their opinions on a twitter-like platform, or post their portfolios, demonstrating their trading operations and returns. Different from those general social networks, almost all the information on Xueqiu is related to stocks, making it a natural data source to collect investors' perceptions, which may be useful for stock market prediction in China.

The literature on stock market prediction was early based on the Efficient Market Hypothesis (EMH) and random walk theory \cite{Fama1965}. However, investors'€™ reactions may not support a random walk model in reality. Behavioral economics has provided plenty of proofs that financial decisions are significantly driven by sentiment. The collective level of optimism or pessimism in society can affect investor decisions \cite{Prechter1999,Nofsinger2005}. Besides, investor perceptions on the relatedness of stocks can also be a potential predictor. Firms may be economically related with one another \cite{King1966,Pindyck1993}. Therefore, there is a probability that one stock's price movement can influence its peer's due to the investment reactions driven by investors'€™ perceptions on such relatedness.


Sentiment and perception are psychological constructs and thus difficult to measure in archive analyses. News articles have been used as a major source for textual content analysis. For example, news articles are employed to analyze public mood~\cite{Li2014}, by which stock price movements can be predicted. However, this type of content has an obvious drawback that news articles directly reflect their authors' sentiment rather than the investors'. Online social platforms have provided us with more direct data and enable opportunities for exploring users' sentiment and perception. In recent studies, it is found in~\cite{JohanBollen2010} that collective mood derived from Twitter feeds improved the prediction accuracy of Dow Jones Industrial Average (DJIA). Facebook's Gross National Happiness (GNH) index is shown to have the ability to predict changes in both daily returns and trading volume in the U.S. stock market~\cite{Karabulut2013}. The predictability of StockTwits (Twitter-like platform specialized on exchanging trading-related opinions) data with respect to stock price behavior is reported in~\cite{Nasseri2015}. 

Most of the existing studies have focused on the U.S. stock market and lacked attention to certain emerging countries such as China, where the stock market is inefficient exhibiting a considerable non-random walk pattern~\cite{Darrat2000}. The China's stock market (also denoted as the A-share market) differs remarkably from other major markets in the structure of investors. Specifically, unlike other major stock markets, which are dominated by institutional investors, retail investors account for a greater percentage in China's market. Importantly, retail investors are more likely to buy rather than sell stocks that catch their attention and thus tend to be influenced by news or other social medias \cite{Barber2008}. Therefore, in this paper, we study the China's stock market based on a unique dataset from a popular Chinese Twitter-like social platform specialized for investors, namely Xueqiu (which means 'snowball' in Chinese), aiming to fill this gap in the literature.

\begin{figure}[h!]
\centering
\includegraphics[width=1.05\linewidth]{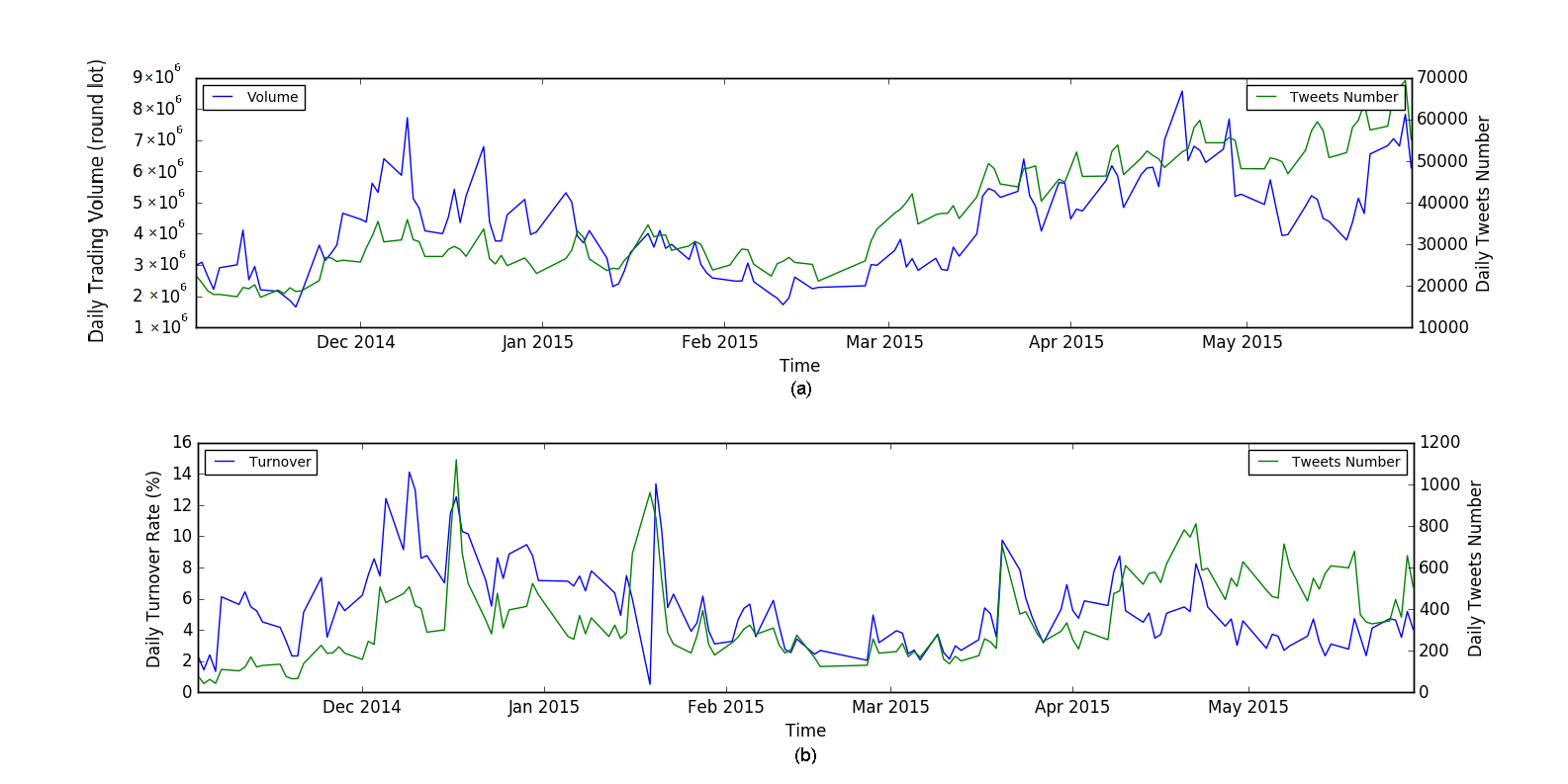}
\caption{Tweets number of Xueqiu vs. trading volume and turnover}
\label{figure1}
\end{figure}

To demonstrate how closely Xueqiu is related to the China's stock market, Figure~\ref{figure1} (a) shows the daily published tweets volume of all stocks on Xueqiu and the daily trading volume of the A-share market from November 2014 to May 2015. It can be observed that the fluctuation trends of these two curves show great synchronicity, especially when high trading volume volatility occurs. When we look at the individual stocks, the synchronicity between the movement of daily tweets volume and the movement of daily turnover rate still holds, as displayed in Figure~\ref{figure1} (b), where one of the most popular stocks in Xueqiu, that is, the CITIC Securities, is taken as an example. On the basis of the tweets from Xueqiu, we analyze features with regard to collective sentiment and perception. The sentiment and perceived stock relatedness are proposed to be formed on the basis of two types of networks extracted from Xueqiu. One is the user network, and the other is the stock network perceived by users. Combined with the network characteristics, the features can exhibit better predictive performance. In contrast to previous works that only study a small subset of the stocks, we evaluate our proposal on all the active stocks (more than 2000) in the A-share market, indicating it's a feasible approach.


In the remainder of the article, we first briefly introduce related research in Section 2. The online social platform Xueqiu and the crawled dataset are described in Section 3. Then, we describe the methodology in Section 4 and present the experiment of predicting stock price movements in Section 5. Finally, the article is concluded in Section 6.

\section{Sentiment, Perception, and Stock Market}

\subsection {Stock Prediction with Historical Price Data} 

Most of previous studies utilize historical stock prices to make predictions with various models~\cite{patra2009computationally,jia2016investigation,Chiang2016,Chong2017}. A Support Vector Machine-based model is proposed by using the selected subset of financial indexes as the weight inputs~\cite{6706743}. A multi-layer perceptron method is proposed for short-term stock prediction in~\cite{6072853}. Multiple techniques of Artificial Neural Network (ANN) in stock market prediction are evaluated in~\cite{7754884}. However, these works only uses the historical price data and ignores the impacts of social media.

\subsection {Sentiment and Stock Price Movement}

A variety of studies have found that financial news can have significant effects on stock price movements~\cite{Cutler1998,Wang2014,Xie2013,Dougal2012,Ahern2015}. Recent studies try to extract events from the news with natural language techniques for event-driven prediction~\cite{Ding2014}~\cite{Ding2015}. News sentiments are measured and the combined effect of Web news and social media on stock markets are studied in~\cite{li2014media}.The investors' sentiments can also be extracted from social networks, media platforms, and blogs. It is reported that social networks such as twitter~\cite{JohanBollen2010}~\cite{Si2014} and Facebook~\cite{Karabulut2013} are important sources of sentiment data. Specialized social networks, such as StockTwits, has also shown its predictive power~\cite{Nasseri2015}. A method to measure the collective hope and fear on each day and analyze the correlation between these indices and the stock market indicators is proposed in~\cite{zhang2011predicting}. A topic-based sentiment time series approach is proposed to predict the market~\cite{si2013exploiting}. This work is extended to further exploit the social relations between stocks from the social media context. A stock network is built with Twitter by co-occurring relationships, and a labeled topic model is employed to jointly model the tweets and the network structure to assign each node and each edge a topic respectively. Then, a lexicon-based sentiment analysis method is used to compute the sentiment score for each node and edge topic. Last, the sentiment time series and price time series are used for prediction~\cite{si2014exploiting}. Financial trend prediction can be boosted with Twitter moods based on deep network models~\cite{huang2016exploiting}. Sentiments and events are integrated with a tensor for stock prediction in~\cite{li2015tensor}.

Overall, most of the prior studies focus on the English social media and U.S. stock market, with little attention paid to China's stock market and China's social media. To predict China's stock market, we are going to conduct analysis on Chinese social media to extract the sentiments. In this work, we choose Xueqiu, a specialized Chinese microblog platform used by millions of investors. Our research will investigate whether sentiments extracted from Xueqiu can be useful for China's stock market prediction.


\subsection {Investor Perceived Stock Relatedness and Stock Price Movement}

Stock correlations are important to understand the behavior of the stock market, and can be measured in various ways. A model of coupled random walks is proposed to model stock correlations, and the walks are coupled via the price change triggered by the price gradients over some underlying network~\cite{PhysRevE.70.026101}. The correlations between stocks are reflected by a stochastic correlation model in~\cite{Chen938934}. Time-series stock correlations are modeled as a mean reverting process, together with a term related to index return~\cite{sepp2011modeling}. In~\cite{preis2012quantifying}, the average correlations among stocks are found to scale linearly with market stress reflected by normalized DJIA index returns on various time scales.

Due to the advances of social media, the human perceived stock relatedness can be captured. Some of the relatednesses is latent or instant, making them important complementaries to tradition classification schemes such as Standard Industry Classification (SIC) scheme. The textual similarity in firms' self-reported business descriptions in their filings is analyzed to infer product market-based peers \cite{Philips2010}~\cite{Philips2014}. Relatedness is defined as a large share of common users on the internet message boards of two companies \cite{DasandSisk2005}, suggesting that stocks associated with each other on message boards may exhibit stronger comovement. The news-based measure of relatedness on investor perceptions of stocks on Twitter is proposed in \cite{Sprenger2011}, which can help delineate meaningful industry groups. A co-search algorithm is applied to Internet traffic at the SEC's EDGAR website \cite{Lee2014}, for identifying economically-related peer firms and for measuring their relative importance. The notion of the semantic stock network from twitter is proposed in \cite{Si2014}, using the topic sentiments from close neighbors of a stock to improve the prediction of the stock market. Stock network models study the correlations of stocks in a graph-based view. Different from the common approaches that measure the pairwise correlations of stocks' historical price series, our approach leverages Xueqiu and identifies the pairwise stocks that are mentioned in one contagion. Our research investigates whether the relatedness extracted from Xueqiu can help to predict the stock price movement.

\section{Data Description}
This section gives details on the mechanism of Xueqiu and the dataset used in this paper. We also conduct data analysis to show the characteristics of Xueqiu.

\subsection{The Mechanism of Xueqiu}

\begin{figure*}[h!]
 \centering
 \setlength{\fboxrule}{3pt}
 \fbox{\includegraphics[width=0.98\linewidth]{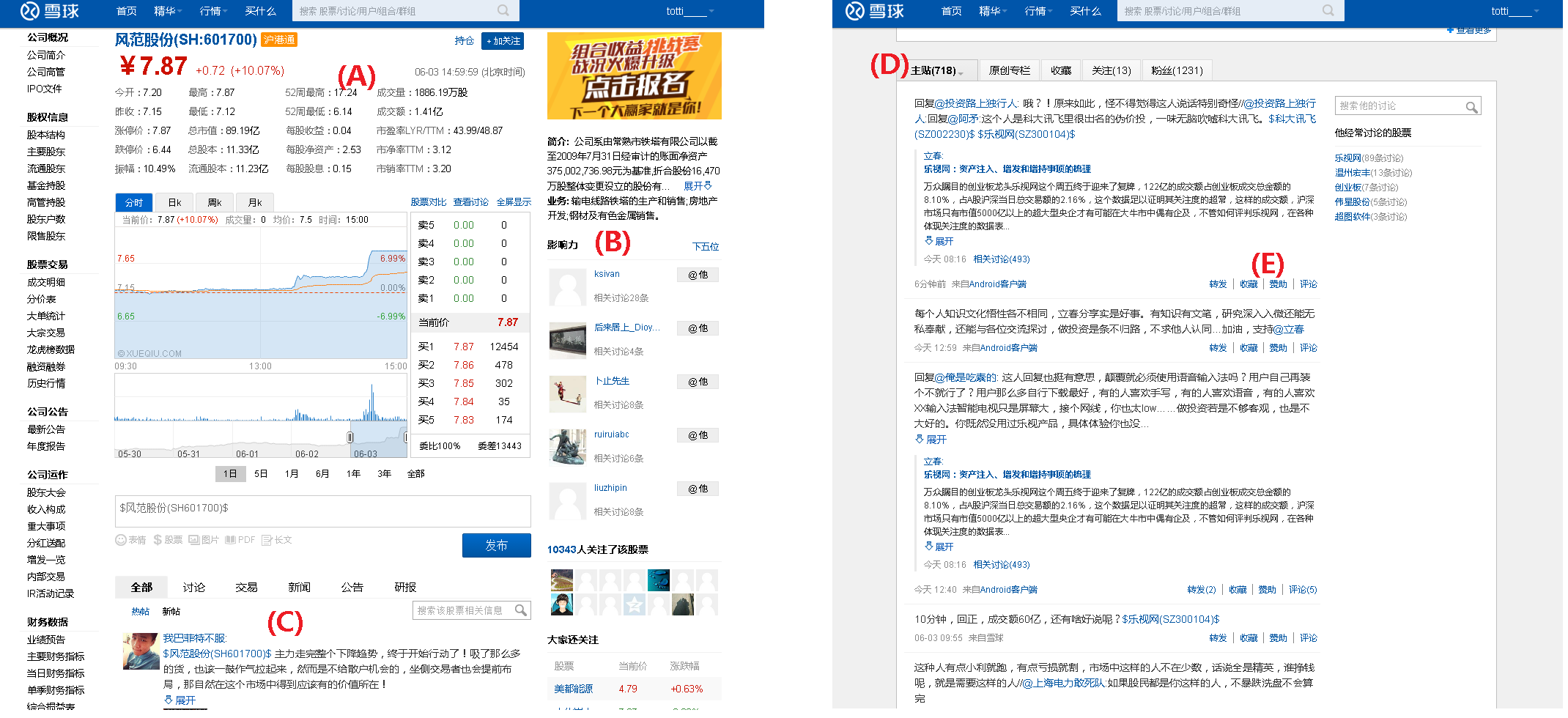}}
 \caption{Xueqiu web pages}
 \label{xueqiu}
\end{figure*}

Xueqiu is a specialized social network for Chinese investors of the stock market. It was established in 2010, and mainly focused on U.S. stock market at first. Since 2014, more and more attention has been paid to A-share market. By the end of 2015, there had been millions of registered users. Xueqiu enables investors to share their opinions on a twitter-like platform, that is, a user can post, reply or repost others' contagions. In addition, each user can follow or be followed by other users, and the number of followers demonstrates his/her authority in some degree. The administrators and official agency of quoted company usually publish authoritative announcements on Xueqiu. In addition to the announcements and opinions, a number of investors post their portfolios, demonstrating their trading operations and returns. Different from general social networks such as twitter or Weibo, almost all the information on Xueqiu is related to stocks, making it a natural data source to collect investors' perceptions.

A typical web page of Xueqiu is shown in Figure~\ref{xueqiu}, which shows the profile of a stock with critical information marked with red letters. The part marked with (A) demonstrates the market indicators on a specific stock, involving the current price, K-line and so on. Part (B) shows the authority users who focus on this stock. Part (C) shows recent tweets talking about this stock. In addition to the stock profile, there are also pages showing user profile. By clicking one user in part (B), we can see the user's information and the published tweets in part (D). Field (E) shows the number of comments and retweets of a tweet.


%

\subsection{Dataset}
We obtain a complete dataset of all users and tweets from December 2010 to May 2015, which consists of 18.39 million tweets from 2,780 stocks (total 2,780 stocks till July 2015) and 2.77 million users. Then we restrict our analysis to the interval from November 2014 to May 2015 for two reasons. First, as some features of the data (e.g., follower graph) keep evolving, we have to choose a relatively short interval with the assumption that such features are stable within this period. Second, the A-share market was very active in this period, resulting in large fluctuations in the market indicators and a lot of discussion tweets on Xueqiu. The dataset we analyze in tis paper involves 6.48 million tweets from 284 thousand active users, and is categorized as users and tweets.


\begin{itemize}
\item \textbf{Users}. For users' information, we crawled user ID, the number of followers, the list of the followers, and the number of the published tweets.

\item \textbf{Tweets}. For tweets, we record not only the content but also the associated attributes, such as the tweet's ID, publishing time, replying and retweeting time. We also record the retweeting behaviors, including the ID of the new tweet and the ID of the user who retweets it.
\end{itemize}

\subsection{Characteristics of the Data}

We begin with the structural analysis of the dataset, and the characteristics observed can help us better understand Xueqiu, and facilitate our prediction task.

\paragraph{\textbf{Distribution of Followers Counts}} \hspace{0pt} 

\begin{figure}[h!]
\centering
\includegraphics[width=0.7\linewidth]{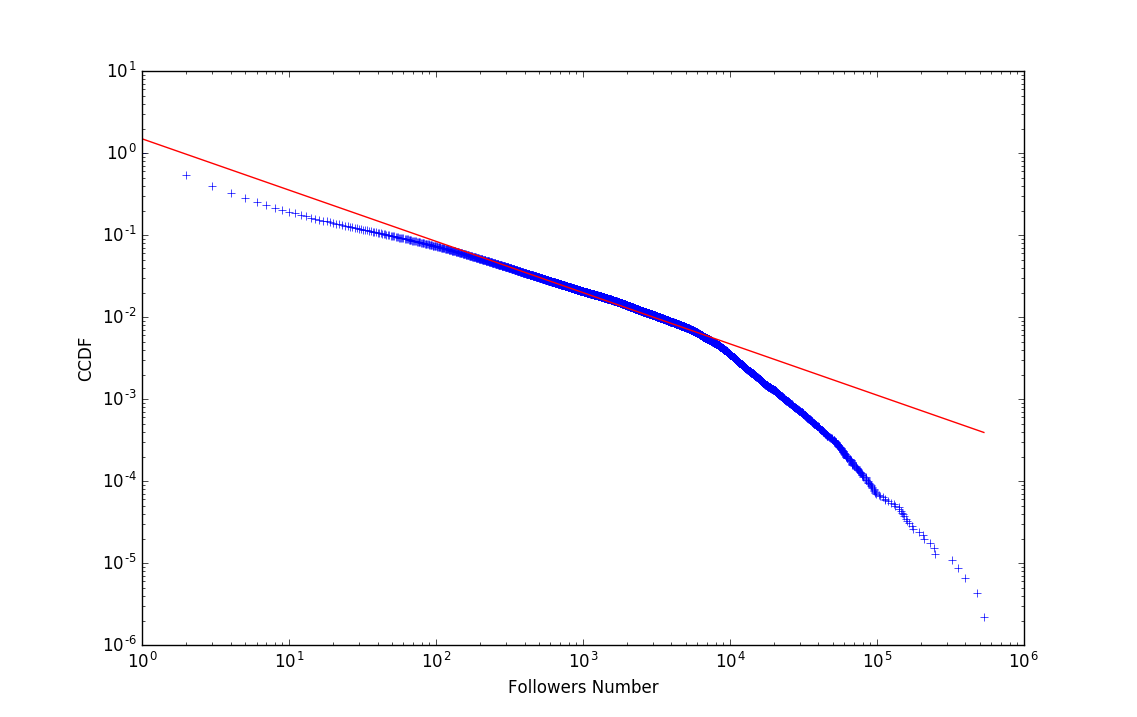}
\caption{Distribution of followers counts}
\label{figure2}
\end{figure}

We first look at the distribution of the follower counts. As shown in Figure~\ref{figure2}, the x-axis represents the number of followers of each user, and the y-axis shows the Complementary Cumulative Distribution Function (CCDF). The blue line shows the results of our dataset, while the red line shows a power law distribution with the exponent of -0.624 and $R^2=0.982$. It can be observed that the distribution curve fits well with the power law distribution when $x \leq 10^4$. The turning point appears at $x=10^4$, and then the blue curve drops quickly. The reason is that only 0.44\% of the total 284 thousand users have more than $10^4$ followers, making it difficult to keep consistent with the other users all through the curve. 

\paragraph{\textbf{Followers vs. Retweeted Counts}}\hspace{0pt} 

When a tweet is retweeted, its influence gets spreading. The retweet count indicates its influence. Generally, a tweet posted by a celebrity can get retweeted easily. We attempt to demonstrate the relations between the number of the followers and the number of the retweets.

\begin{figure}[h!]
\centering
\includegraphics[width=0.7\linewidth]{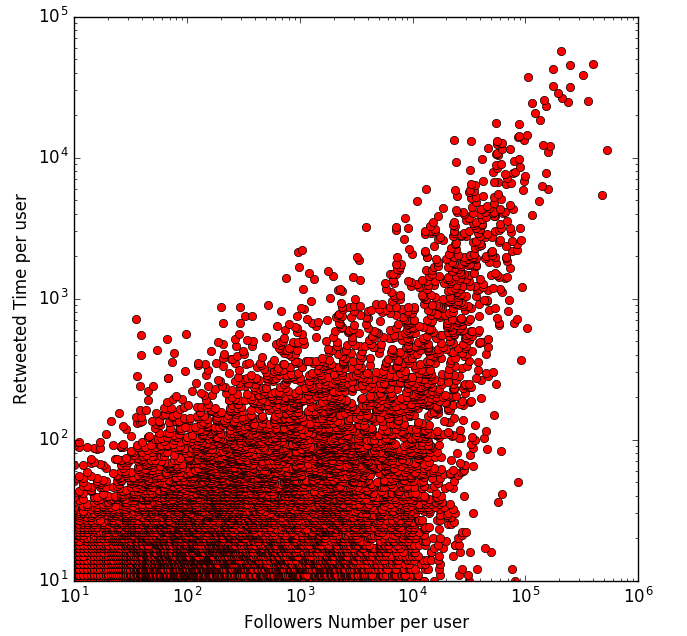}
\caption{Followers counts vs. retweeted counts}
\label{figure3}
\end{figure}

The scatter diagram is shown in Figure~\ref{figure3}, where the x-axis represents the number of followers and the y-axis stands for its retweet number.  It can be observed that when the number of followers $x$ exceeds $1.2\times10^4$, the number of retweets is larger than $1\times10^3$, indicating that tweets published by a celebrity whose follower number is large enough (larger than $10^4$) can get retweeted much more easily. Moreover, as the increasing of the followers, the number of retweets grows linearly, especially when  $x>10^4$. 

%
%

\subsection{Sentiments vs. A-Share Indicators}

%
In order to investigate the correlations between sentiments of the tweets and the stock prices, we first extract the sentiments from tweets, and use Naive Bayes Algorithm to infer the sentiments. Tweets are classified into three categories: negative, positive and neutral. Negative and positive tweets are applied to construct the sentiment index at some day $i$, which is defined as

$$S_i=0.5-\frac{\frac{p_i}{\sum_{i=1}^N p_i}}{\frac{p_i}{\sum_{i=1}^N p_i}+\frac{n_i}{\sum_{i=1}^N n_i}}$$

where $N$ is the number of dates, and $p_i$ and $n_i$ are positive and negative tweets numbers at day $i$, respectively.

\begin{figure}[h]
\centering
\includegraphics[width=1.1\linewidth]{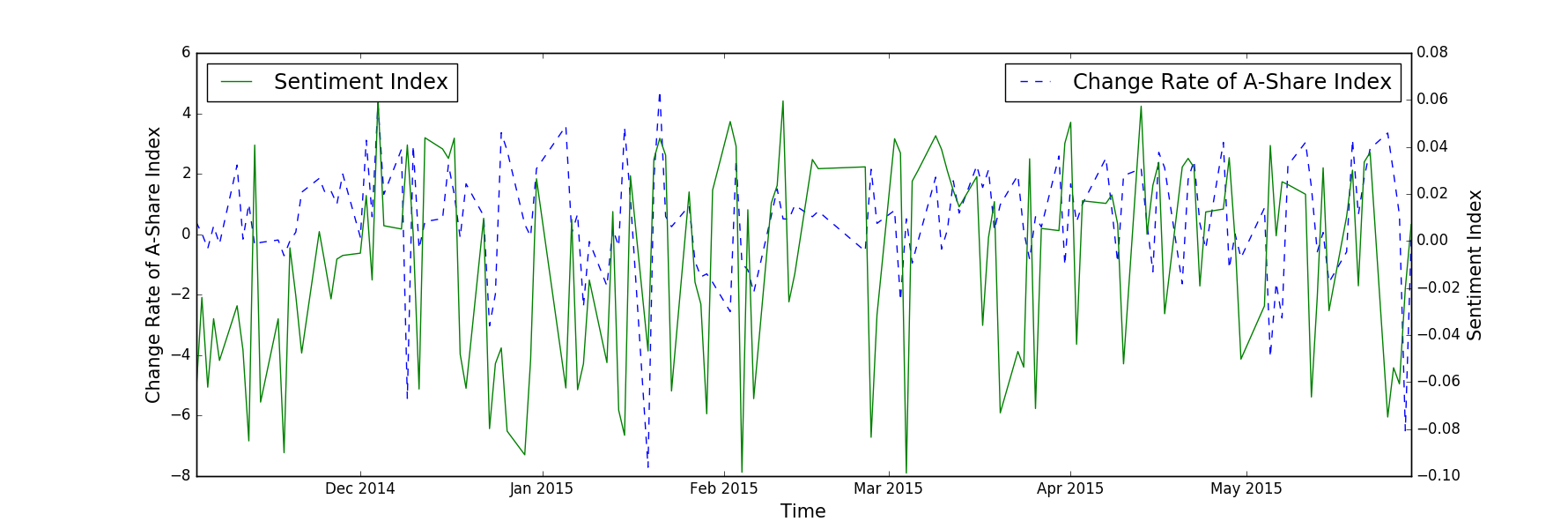}
\caption{Change rate of A-Share index vs. sentiment index}
\label{figure5}
\end{figure}

In Fig.~\ref{figure5}, the solid (green) curve represents the sentiment index from Dec. 2014 to May. 2015, and the dashed (blue) curve stands for the change rate of A-share Index, i.e., the Shanghai Stock Exchange Composite Index, in the same time interval. If these two curves coevolve, it indicates the sentiments presented by tweets on Xueqiu are correlated with A-share Index. It is reasonable that the positive emotion usually goes with the rise of the stock price, and vice versa. It can be observed that in a lot of time intervals, these two curves show similar fluctuation trends, especially at the peak or bottoms of the A-share index curve. For example, on Jan. 19th, 2015, the A-share index dropped by 7.7\%, and the sentiment index also dropped a lot, indicating the strong negative emotions of the investors.

%

\subsection{User Perceived Stock Relatedness}

The common method to obtain the stock correlations is to use the standard industry classification scheme or historical price series. However, in this paper, we extract the user perceived relatedness through Xueqiu. The advantage is that, in addition to the explicit and static relatedness, we can also obtain the latent or instant correlations, e.g., the correlated stocks which are driven by the same event but not affiliated with the same industry.
%

\begin{table}
\centering
\caption{Top 5 co-occurrence statistics}
\vspace{3pt}
\label{tab:cooccu}
\begin{adjustbox}{max width=1\columnwidth}
\begin{tabular}{lllll}
\hline
&China Merchants Bank & LeTV & ChinaNetCenter & Kweichow Moutai \\\hline
1 &  Industrial Bank & EastMoney & LeTV & China Merchants Bank \\\hline
2 & SPD Bank & Hithink RoyalFlush & Ourpalm & Luzhou Laojiao \\\hline
3 & Minsheng Bank & CITIC Securitie & Kweichow Moutai & Ping An \\\hline
4 & Ping An & Siasun Robot & Sinnet &  Gree \\\hline
5 & CITIC Securities & Huayi Brothers & Siasun Robot & SPD Bank \\\hline
\end{tabular}
\end{adjustbox}
\end{table}

To obtain such correlations, we collect all the pairwise stocks mentioned by the same tweet. For preprocessing, we removed tweets mentioning more than five continuous stock tickers as such tweets usually do not convey much meaning for our task. Table~\ref{tab:cooccu} shows the top five most frequent stocks jointly mentioned with China Merchants Bank, LeTV, ChinaNetCenter and Kweichow Moutai respectively. It can be observed that the top 5 stocks related to China Merchants Bank are all financial companies, the top three are banks that have similar sizes as China Merchants Bank, and the forth is Ping An, a comprehensive financial company involving the Ping An Bank. Citic Securities, the largest securities company in China, takes the fifth place. For the stock LeTV, the correlated stocks are diverse. LeTV is a company whose major products are smart TVs and video services, while EastMoney is a website providing financial news and data. Though they are not in the same industry, they are treated as representative companies in China Growth Enterprise Board by investors, and thus co-occurrence frequently. For ChinaNetCenter and Kweichow Moutai, their most correlated stocks are also not restricted to the same industry. Thus, it can be summarized that the user perceived relatedness from Xueqiu can capture implicit correlations which are difficult to observe by previous methods. Correlations may result in coevolving in stock prices, and our work is to investigate whether such correlations extracted from Xueqiu is effective for our prediction task.

\section{Prediction of Stock Price Movement}

In this section, we model the prediction of stock price movement as a binary classification problem. Then we discuss how to extract features from three different types of information sources. After that, we evaluate the classification model to verify the effectiveness of the information from Xueqiu.

\subsection{Problem Formulation}

The movement of stock price only happens during trading days, so we define one single trading day as the time granularity of our prediction model. A trading day is defined from the close time (i.e. 3:00pm) of the last day to the close time of today. We would predict whether the close price of today is increased or decreased compared to the close price of the last day. Given a target stock $s_i$, a series of its continuous valid trading days constitute the trading day vector $\vec{T_i}=( t_1^i,t_2^i,t_3^i,...,t_n^i )$, where $n$ is the number of the trading days in $\vec{T_i}$, determined by the range of the dataset. Note that different stocks would have different trading day vectors. 

For some trading day $t_j^i$, we define feature vector $\vec{x_j^i}$,  consisting of features extracted for stock $s_i$ at trading day $t_j^i$. The feature vector is also the input of the prediction model. Formally, given the stock $s_i$ and its feature vector $\vec{x_j^i}$, the stock price movement prediction problem is modeled as:

\begin{equation}
 y_j^i = f(\vec{x_j^i})=
 \left\{
 \begin{array}{ll}
 1, & if\ price\ of\ s_i\ increases\ on\ t_{j+1}^i\\
 0, & otherwise
 \end{array}
 \right.
\end{equation}

Where $y_j^i$ is the result of the prediction function $f(\vec{x_j^i})$, denoting the price movement direction of stock $s_i$ at the next trading day $t_{j+1}^i$.

\subsection{Feature Extraction}
Motivated by the data analysis in Section 3, we explore rich knowledge from Xueqiu and stock market to constitute the input feature vector $\vec{x_j^i}$, and categorize the features into three types, the stock specific features, the sentiment feature and the stock relatedness feature.

\paragraph{\textbf{Stock Specific Features}}\hspace{0pt} 

The common information used for stock prediction is the firm-specific factors, as well as the historical and time-series prices used for technical analysis~\cite{TaylorandXu1997}~\cite{Taylor2007}. We select some key characteristics of a stock which show the predictive ability to some degree in previous literature~\cite{Li2014}~\cite{fama1992cross}: stock price, trading volume, turnover and price-to-earnings (P/E) ratio. Note that the absolute value of stock price and trading volume would have huge difference between different stocks, so we use the change rate instead. Besides, not only the daily change but also the change of 5-days-moving-average is involved.


\paragraph{\textbf{Sentiment Features}}\hspace{0pt}

In this study, we derive the sentiment index for each stock at each trading day. Firstly, for all the tweets in the dataset, we classify them into three categories: positive, neutral and negative, and only the positive and negative tweets are used to derive the sentiment feature. Counting the number of tweets in the positive and negative categories is an intuitive way to measure how strong each sentiment is. However, this counting method implies that each tweet is treated equally and thus has the same weight. In fact, it's clear that different tweets might have different influences due to different authorities of the users. Thus, it is reasonable to take the user's authority as the weight for each tweet. 

\begin{figure}[h]
\centering
 \setlength{\fboxrule}{3pt}
\includegraphics[width=8cm]{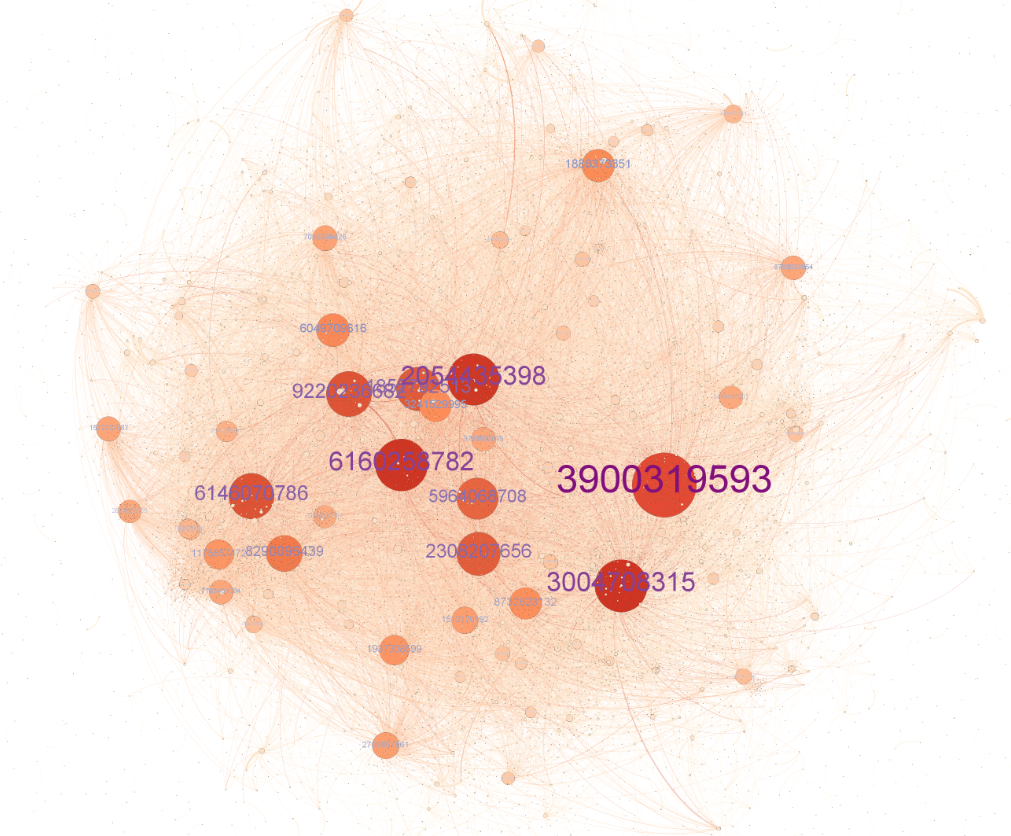}
\caption{A sample of user network}
\label{figure6}
\end{figure}

Given a user network extracted from Xueqiu, PageRank is a natural method to weight each user. To derive the PageRank score, we first construct the user network from the dataset. Note that different from the static friendship links in the social network, the user network constructed here is a dynamic forwarding network. Specifically, as the users publish tweets or forward others' tweets on Xueqiu, a user forwarding network can be constructed. Figure~\ref{figure6} shows a sample of the user network on May 29th, 2015. 
In this network, each node stands for a user marked with its user ID, and the edge stands for the forwarding behavior between the two users. There are totally 141 trading days of the A-share market in our dataset, so 141 user networks are constructed. For each network, we calculate PageRank value of each vertex. As shown in Figure~\ref{figure6}, the bigger a node is, the larger the user's PageRank is.

%
%
For a user $v_t$ in the directed user network, given a user set $U$ with $K$ users (denoted as nodes ${u_1,u_2,...,u_K}$) have forwarded $u_t$'s tweet, the PageRank value of $u_t$ can be calculated as


\begin{equation}
PR(u_t) = \sum_{i=1}^{K} \frac{PR(u_i)}{L(u_i)}
\end{equation}
where $L(u_i)$ is the number of outbound links from $u_i$. After that, the weight of each tweet $x$ is the weight of the user $u(x)$ who has published it, and then the weighted count is

\begin{equation}
\begin{split}
PositiveCount=\sum_x PR(u(x)) \\
NegativeCount=\sum_y PR(u(y))
\end{split}
\end{equation}
where $x$ and $y$ denote the positive tweet and negative tweet respectively.

For a given stock $s_i$ and some trading day $t_j^i$, we first calculate its positive count and negative count, and then combine them into one sentiment score, denoted as $SI_j^i$, that is

\begin{equation}
SC_j^i = \frac{PositiveCount_j^i}{PositiveCount_j^i+NegativeCount_j^i}
\end{equation}

Obviously, $SC_j^i \in [0,1]$, and the larger $SC_j^i$ is, the more positive the overall emotion is. $SC_j^i$ is used as the sentiment feature for our prediction model.





%

\paragraph{\textbf{Stock Relatedness Features}}\hspace{0pt} 

User-perceived relatedness among stocks is another knowledge that could be obtained from Xueqiu~\cite{Arai2015}. The intuition is that stocks with strong correlations may demonstrate comovements on prices. In our work, stocks are regarded as correlated stocks if they are jointly mentioned by a tweet. Formally, we define the stock network as an undirected graph $G=\{V, E\}$. The node set $V$ comprises of stocks, and $e_{u,v} \in E$ stands for the edge between stock nodes $u$ and $v$ and the edge weight is the number of co-occurrences in the last 3 days.
As this correlation is time-sensitive~\cite{Wichard2004}, we construct 141 stock networks for 141 trading days. 

%


Specifically, for a given stock $s_i$ and the trading day $j$, let $r_j^{i,k}$ denote the weight of the edge between stock $s_i$ and $s_k$ at day $j$. To make the correlation more specific and meaningful, we filter the non-informative edge with $r_j^{i,k} < 2 $  (except $r_j^{i,i}$). Note that $r_j^{i,i}=1$. For any two stocks (namely $s_m$ and $s_n$) which are not connected in the stock network,  $r_j^{m,n} = 0$.


Then given $r_j^{i,k}$ as the weight, we can combine it with a stock specific feature $f_k$ of the stock $s_k$ to obtain the relatedness feature at day $j$, that is

\begin{equation}
corr(f)_j^i = \frac{\sum_{k=1}^{N} r_j^{i,k}f_k}{\sum_{k=1}^{N} r_j^{i,k}}
\end{equation}
where $N$ is the number of stocks in the dataset and $f_k$ is a stock specific feature of the stock $s_k$. Take turnover rate and  stock price change rate as examples, we can obtain

\begin{equation}
\begin{split}
corr(turnover)_j^i = \frac{\sum_{k=1}^{N} r_j^{i,k}turnover_k}{\sum_{k=1}^{N} r_j^{i,k}}\\
corr(price\_change)_j^i = \frac{\sum_{k=1}^{N} r_j^{i,k}price\_change_k}{\sum_{k=1}^{N} r_j^{i,k}}
\end{split}
\end{equation}

\subsection{Prediction Methods}

Given the feature vector, we then apply statistical learning methods to obtain the prediction results. Specifically, given a training set of $n$ points with the form $(\vec{x_1},y_1),...,(\vec{x_n},y_n)$, where $y_i$ is either +1 or -1. The Class +1 denotes that the stock price will increase, while the Class -1 means the stock price will decrease. $\vec{x_i}$ is a vector for a specific stock on a certain day containing the features applied to train the model. 

To obtain the prediction results, we consider both the Support Vector Machine (SVM)~\cite{Boser1992} and the Multilayer Perceptron (MLP)~\cite{hinton1987learning,rumelhart1988parallel} algorithms. Most previous works use linear models to predict the stock market~\cite{xie2013semantic,luss2015predicting,kogan2009predicting}. However, the relationship between the features and the stock price movements may be more complex than linear. Thus, we use RBF-kernel instead of the linear kernel in SVM, and the results also show that using RBF kernel is better than using linear kernel. In addition, we also exploit the MLP model to learn the hidden and complex relationships. MLP is a feedforward artificial neural network model that maps sets of input data onto a set of appropriate outputs. An MLP consists of multiple layers of nodes in a directed graph, with each layer fully connected to the next one. The structure of the model in our work is using one hidden layer and using sigmoid as the activation function. The standard back-propagation algorithm is used for supervised training of the neural network.

The process of feature extraction and prediction is shown in Algorithm 1.

\floatname{algorithm}{Algorithm}  
\renewcommand{\algorithmicrequire}{\textbf{Input:}}  
\renewcommand{\algorithmicensure}{\textbf{Output:}}  
    \begin{algorithm}  
        \caption{Process of Feature Extraction and Prediction} \label{a:1}
        \begin{algorithmic} 
            \Require Users $U$ and tweets $X$ from Xueqiu, firm-specific factors $F$ of stock $s_i$ at trading day $t^i_j$ 
            \Ensure Stock movement $y^i_j$ at next trading day $t^i_{j+1}$ 
            \Function {SpecificFeature}{$F$,$s^i_j$}  
                \State Extracting firm-specific features: $f^i_j\gets F$ for stock $s^i_j$ at $t^i_j$;
                \State \Return{$f^i_j$}  
            \EndFunction  
            \\
            \Function{SentimentScore}{$X$,$U$,$s_i$}  
                \State Counting the number of tweets in the positive category (i.e., $x$) and negative category (i.e., $y$) respectively;
                \State Constructing the user forwarding network;
                \State PageRank value for user $u_t$: $PR(v_t)\gets \sum{\frac{PR(u_i)}{L(u_i)}}$;
                \State Positive weighted count: $PositiveCount\gets \sum_x{PR(u(x))}$ 
			\State Negative weighted count: $negativeCount\gets \sum_y{PR(u(y))}$
                \State Sentiment Score: $SC^i_j\gets \frac{PositiveCount^i_j}{PositiveCount^i_j+NegativeCount^i_j}$;
                \State \Return{$SC^i_j$}  
            \EndFunction  
		\\
		 \Function{RelatednessFeature}{$f^i_j$,$X$,$s_i$}  
                \State Constructing the stock network: $G=\{V,E\}$;
                \State Correlation between stock $s_i$ and $s_k$: $r^{i,k}_j\gets \frac{\sum^{N}_{k=1}{r^{i,k}_j f_k}}{\sum^{N}_{k=1}{r^{i,k}_j}} $;
                \State Relatedness feature: $corr(f_k)^i_j$, $f_k$ is a specific feature of stock $k$;
                \State \Return{$corr(f)^i_j$}  
            \EndFunction  
		\\
		 \Function{Prediction}{$f^i_j$, $SC^i_j$, $corr(f)^i_j$}  
                \State Combining features into a vector: $\vec{x^i_j}\gets \{f^i_j,SC^i_j,corr(f)^i_j\}$;
                \State Predicting stock movement: $y^i_j\gets SVM(\vec{x^i_j})$ (or $y^i_j\gets MLP(\vec{x^i_j})$);
                \State \Return{$y^i_j$}  
            \EndFunction  
                   \end{algorithmic}  
    \end{algorithm}

\section{Experiments}

In this section, we conduct experiments to evaluate the effectiveness of using knowledge from Xueqiu to predict stock price movements. 
\subsection{Experimental Setup}
We select the Xueqiu data from Nov. 2014 to May 2015. The target stocks are selected from all the stocks in the A-share market satisfying two requirements: (1) there are more than 10 trading days for that stock during this time; (2) the number of tweets about that stock is more than 10 per day. The spam contagions in Xueqiu may lead to large noises in our analysis and prediction task. To detect the spams, we have determined ten features (including the percentage of digits in the contagion, the number of followers of the user, etc.), and use logistic regression to identify them. Then we extract Xueqiu related features (sentiment features and the stock relatedness features) for each stock. The stock-specific features are extracted from the historical information obtained through TuShare~\footnote{http://tushare.org/}. The sentiment of each tweet from Xueqiu is classified by SnowNLP~\footnote{https://github.com/isnowfy/snownlp}, an open-source Chinese text processing toolkit. Finally, we get about 35.7K valid test samples from our dataset.

We use the SVM (with RBF-kernel) and MLP as the prediction models. The samples in the last month would be used as the training set to predict the stock price movements for each trading day in the following month. For example, when the samples in Nov. 2014 are used as the training set, the trading days in Dec. 2014 are the corresponding testing set. The prediction is evaluated through two commonly used metrics: classification accuracy (ACC) and Area Under ROC Curve (AUC). ACC is very sensitive to the data skew. When a class has an overwhelmingly high frequency, the accuracy can be high using a classifier that makes the prediction on the majority class. Thus, we also use AUC to avoid the bias due to data skew. Though our data is not severely skewed, we also use AUC for comparison. After conducting predictions on all these testing sets, we aggregate all results of AUC and ACC into an overall output.

\subsection{Prediction Results}

\begin{figure}[h]
\centering
\includegraphics[width=0.7\linewidth]{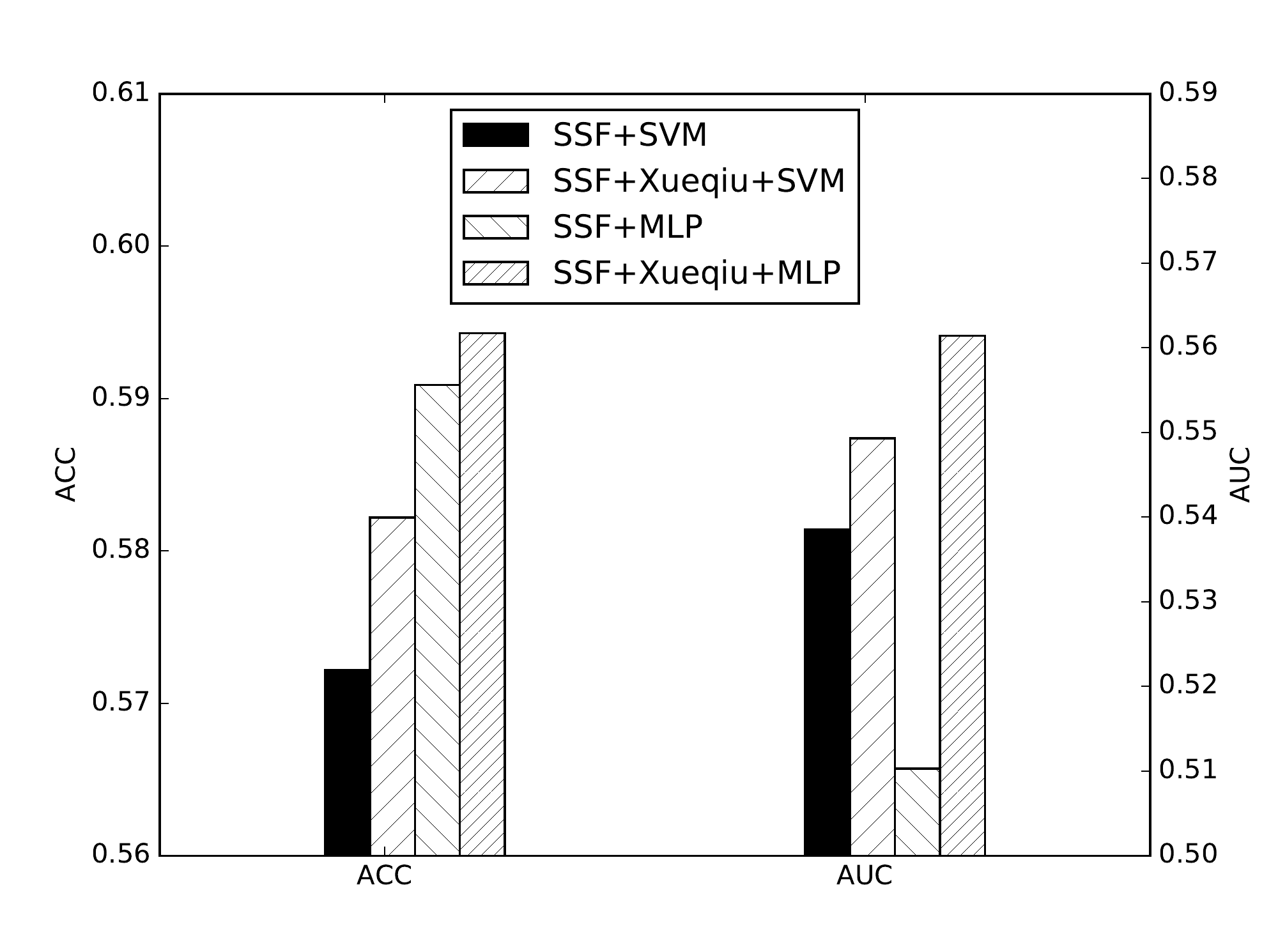}
\caption{Prediction results using SVM and MLP with only stock specific features (SSF) v.s. with both SSF and Xueqiu features}
\label{figure7}
\end{figure}

According to previous studies~\cite{Li2014}~\cite{fama1992cross}, the stock specific features are vital for stock prediction, so using prediction methods (i.e. SVM and MLP) with only stock specific features is adopted as our baselines. To verify whether the knowledge extracted from Xueqiu is effective for stock prediction, the prediction methods with stock specific features as well as the Xueqiu features (i.e., sentiment features and stock relatedness features) are evaluated against the baselines. The results are shown in Figure~\ref{figure7}. It can be observed that given the same prediction model (SVM or MLP), the method involving Xueqiu features achieves consistently better performance than only involving the stock specific features over both ACC and AUC metrics. This confirms that the investors' perceptions extracted from Xueqiu can assist in stock prediction. This also demonstrates that the Chinese social media can have reflected the investors' opinions and behaviors in China's stock market. In addition, the MLP model achieves better performance than the SVM model, partly by effective learning hidden relationships between the features and the price movements. Based on the above analysis, we can observe that both features and algorithms can have impacts on performance. The prediction errors may come from not capturing sufficient and effective features or from not using suitable algorithms. In addition, how we choose the training samples and testing samples may also be a source of errors.

%
%

\subsection{Feature Importance Analysis}

\begin{figure}
\centering
\includegraphics[width=0.7\linewidth]{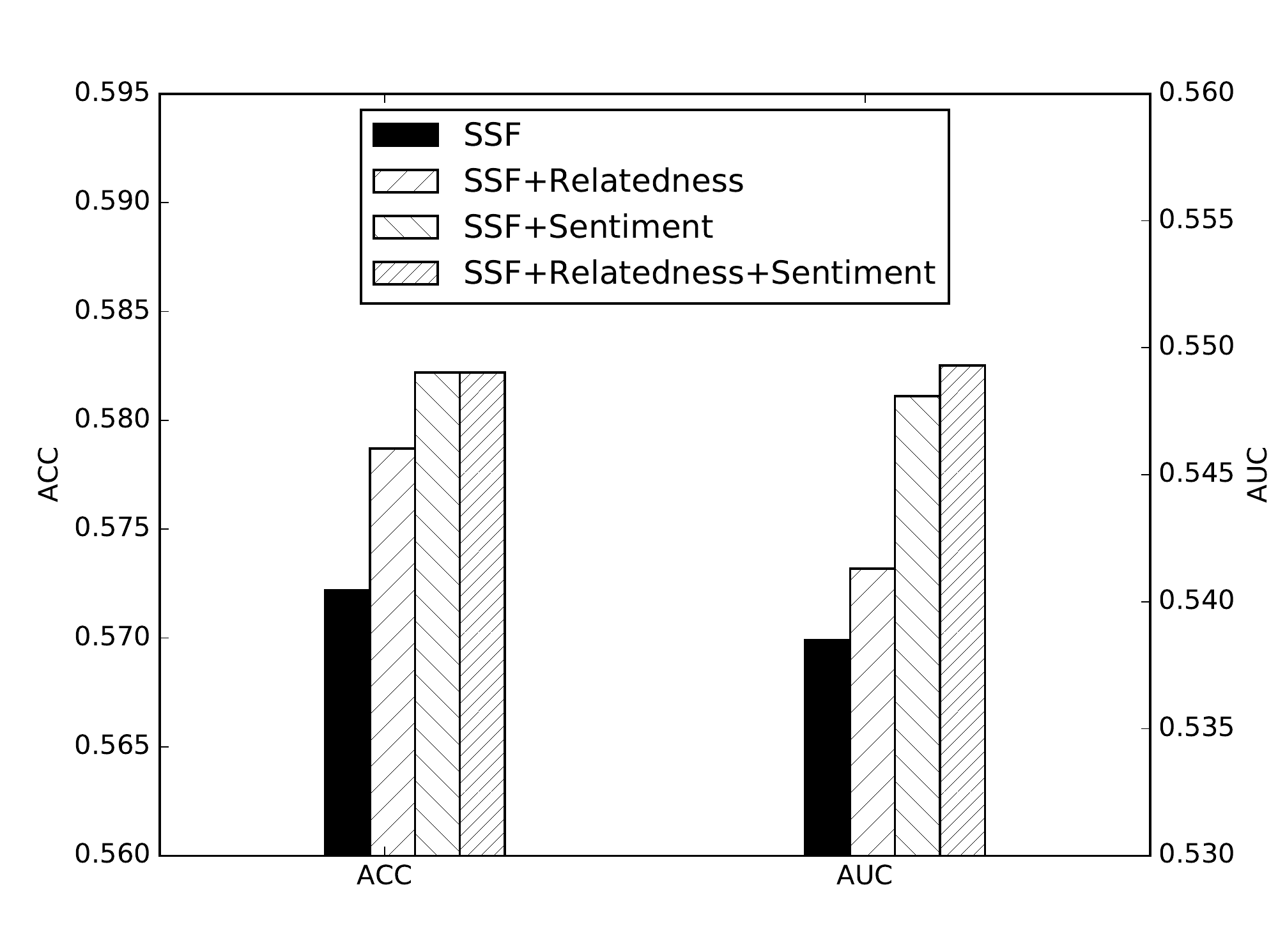}
\caption{Prediction results trained on different combinations of features with SVM}
\label{figure8}
\end{figure}

Feature important analysis studies how important the various features are in the prediction task. From a macroscopic view, we first study the importance of the features derived from different types of knowledge. Figure~\ref{figure8} shows the prediction results with different groups of features. Not surprisingly, the stock specific features are very useful for stock prediction. Using stock specific features alone can achieve 0.57 in ACC and 0.54 in AUC. Both the sentiment features and the stock relatedness features are helpful, and the sentiment features play more important roles than the stock relatedness features. When we put all features together, the prediction result in terms of ACC remain the same as that with both sentiment and stock specific feature, but the results in terms of AUC can be further improved. The reason is that the imbalance in the dataset is not considered in ACC. The improvement in AUC indicates that the addition of stock relatedness feature can improve the discriminative power of our model.

\begin{table}
\centering
\caption{Top-10 features}
\label{table2}
\begin{adjustbox}{max width=1\columnwidth}
\begin{tabular}{l}
\hline\noalign{\smallskip}
Features \\
\noalign{\smallskip}
\hline
\noalign{\smallskip}
Stock Price Change Rate \\
Stock MA5 Value \\
Stock Turnover Change Rate \\
Stock Trading Volume Change Rate \\
Sentiment Score \\
Stock 5-Day Moving Average of Trading Volume \\
Sentiment Tweets Count (Positive) \\
Correlation Stocks Weighted Average (MA5) \\
Sentiment Tweets Count (Negative) \\
Sentiment Tweets Count (Neutral)  \\
\hline
\end{tabular}
\end{adjustbox}
\end{table}

We then study the importance of features from a microcosmic point of view. To evaluate how the features contribute to the prediction results, we use the random forests model to obtain the rank of the importance of the features~\cite{Robin2010}, which is shown in Table~\ref{table2}. It is clear that stock specific features are the most influential feature type in the model, as the top 4 features all belong to it. While additional features are taken into consideration, sentiment features are more important than the stock relatedness features, which is coherent with the feature analysis presented above in Fig.~\ref{figure8}. Among sentiment features, Sentiment Score is the most critical one containing not only the contents of the tweets but also the structure of the user network. Correlation Stocks Weighted Average (MA5) is in the top 10 features, indicating it is also useful for our task.


\subsection{Summary}

We summarize the experimental results by the following observations:

(1) \emph{Knowledge extracted from Xueqiu is useful for stock prediction in China's stock market. }The results show that by exploiting the knowledge from Xueqiu, the prediction results can be consistently improved in terms of both ACC and AUC. Previous studies show the effectiveness of sentiments for stock prediction in the U.S. stock market. Our research confirms that the sentiments extracted from Xueqiu are also effective for China's stock market. In addition, we also observe the effectiveness of user perceived stock relatedness in stock prediction. 

(2) \emph{The most important features are stock specific features.} Both the prediction results with only stock specific features and the feature importance analysis show that the stock specific features are the most important ones for our task.

(3)\emph{The sentiment feature is more important than the stock relatedness features.}  One possible reason is that the user perceived stock relatedness is more sparse than the sentiments. Only a few stocks are mentioned jointly with other stocks in the tweets, especially considering our stock network is time-sensitive, i.e., only the co-occurrences in the last 3 days are taken into consideration. Despite the sparseness, the prediction results are pretty good for almost the full set of stocks in the A-share market .

\section{Conclusions and Future Work}

In this study, we have studied a unique social network, namely Xueqiu, where retail investors'€™ tweets can be employed to extract features such as sentiment and perceived stock relatedness. Further, we adopt SVM model and MLP model to predict the stock price movements in the China's market. The results show that the predictive performance can be improved by including the features of sentiment and perceived stock relatedness. The study contributes to both social network analysis and the behavioral economics literature, by providing a deeper understanding of the investors' perceptions through the social network.

Potential avenues of future work include a deeper study on the different measures of the construct of our interest. It would also be interesting to conduct a more comprehensive analysis on the time-series features together with the temporal models such as Long Short-Term Memory (LSTM) network~\cite{hochreiter1997long} .

\section{Acknowledgement}
This work was supported in part by State Key Development Program of Basic Research of China (No. 2013CB329604), the National Key Research and
Development Program of China (No. 2016QY03D0605), the Natural Science Foundation of China (No. 61300014, 61372191, 61472263), the Project on the Integration of Industry, Education and Research of Guangdong Province (No. 2016B090921001), and DongGuan Innovative Research Team Program (No. 201636000100038).

\end{document}